# 改革开放以来中国城乡居民消费与经济增长的关系
## ——基于计量经济学模型的实证分析

易志珩

长沙理工大学　国际工学院　长沙　410114
Email：zhiheng.yi@stu.csust.edu.cn

**摘　要**：改革开放 40 余年来，中国经济发展取得了举世瞩目的成就，其中包括居民消费在内的消费活动功不可没。消费活动是经济活动的终点，因为其他经济活动的最终目的是满足消费需求；消费活动又是经济活动的起点，因为消费能带动经济社会发展。选取改革开放以来 40 余年的经济数据，通过建立向量自回归（VAR）模型和向量误差修正（VEC）模型，分析城乡居民消费水平和城乡居民总消费对经济增长的影响。得出结论：城镇居民消费和农村居民消费的提升都能导致国内生产总值的提高，且从长期来看，城镇居民消费对经济增长的促进能力比农村居民消费更强。根据这一结论分析原因，并提出政策建议。
**关键词**：消费经济学，计量经济学，实证分析，居民消费，经济发展

# The Relationship between Consumption and Economic Growth of Chinese Urban and Rural Residents since Reform and Opening-up -- An Empirical Analysis Based on Econometrics Models

Yi Zhiheng

International College of Engineering, Changsha University of Science & Technology, 410114, Changsha, China P.R.
Email：zhiheng.yi@stu.csust.edu.cn

**Abstract**: Since Reform and Opening-up 40 years ago, China has made remarkable achievements in economic fields. And consumption activities, including household consumption, have played an important role in it. Consumer activity is the end of economic activity, because the ultimate aim of other economic activities is to meet consumer demand; consumer activity is the starting point of economic activity, because consumption can drive economic and social development. This paper selects the economic data of more than 40 years Reform and Opening-up, and establishes the vector autoregressive (VAR) model and vector error correction (VEC) model, analyzing the influence of consumption level and total consumption of urban and rural residents on economic growth. The conclusion is that the increase of urban consumption and rural consumption can lead to the increase of GDP, and in the long run, urban consumption can promote economic growth more than rural consumption. According to the conclusion, we analyze the reasons and puts forward some policy suggestions.

# 1 引言（Introduction）

消费对于国民经济发展有着重要的作用。改革开放 40 余年来，中国经济发展成就举世瞩目，其中包括居民消费在内的消费活动功不可没。消费活动是经济活动的终点，因为其他经济活动的最终目的是满足消费需求；消费活动又是经济活动的起点，因为消费能带动经济社会发展[1]。中国著名经济学家尹世杰曾研究过居民消费结构与生活水平提高的关系[2]；潘文轩曾根据西方收入假说，得出了中国城乡居民消费存在差异性，政策制定时应根据城乡消费差异来制定的结论[3]。经济增长反映了各国经济与上年相比增长的幅度以及速度，可以用来衡量一个国家的总体经济发展实力。一般来说，消费由居民消费和政府消费两部分组成，其中居民消费在消费中占主导作用，因此，研究居民消费和经济增长的关系具有较大的价值。

# 2 居民消费与经济增长（Household Consumption and Economic Growth）

消费是社会再生产过程中的最终环节，指人们利用社会产品或者社会服务以满足各种需要的过程，主要包括生产消费和个人消费。其中，生产消费是指在生产过程中对劳动和生产资料的消耗，个人消费则是个人利用社会产品或者社会服务以满足需要的过程，这种消费对生产力的恢复与促进意义重大。如果没有特殊说明，后文中提及的消费均是指个人消费。

消费水平能够有效反映社会成员对物质文化需求的满足程度。根据凯恩斯经济理论及边际消费倾向效应可知，消费水平会受收入水平的影响。收入水平的提高会引起消费水平的提高，但是消费水平提高的增速会低于收入水平提高的增速，引起消费水平提高的滞后，从而导致需求的减少。

中国经济发展可以分为两个阶段。第一个阶段是计划经济阶段，具体时间是 1949 年新中国成立后至 1978 年改革开放前；第二个阶段是中国特色社会主义市场经济阶段，具体时间是 1978 年改革开放后至今。

新中国成立初期，抗日战争及内战结束不久。中国的经济体制在战争时期受到了极大的破坏，同时大量黄金的流失进一步恶化了经济发展情况。中国政府不得不采取各种措施恢复经济、积累财富。所实行的低消费经济政策实现了经济的初步发展，为改革开放奠定了经济基础。

1978 年，改革开放政策得以实施，对中国的人口红利、政策红利实现了有效利用，通过投资来引入资本、扩大生产。同时引入了市场经济来逐步取代实行已久的计划经济。此后，消费得以放开，市场的运作抵消了竞争带来的诸多问题，经济总体呈现良性发展态势。在 2008 年的金融危机中，中国经济受到的影响小于欧美国家。

居民消费和经济增长的关系表现在两个方面。

一方面，经济增长对居民消费的推动作用主要表现在两个方面。首先是消费数量的提升。经济增长可以为居民带来更多收入，而收入的增加会直接促进消费。在宏观经济学领域，专家学者普遍认为收入是影响消费最重要的因素之一，收入对消费有着至关重要的作用。另一方面是消费质量的提高。近十年来，中国居民的生活迈出了从小康到富足的关键一步，尤其是在脱贫攻坚完成和乡村振兴战略实施的背景下，消费质量明显提高。经济的发展会促使企业研发生产更加优质的产品，居民的消费质量因此提高。

另一方面，居民消费可以直接和间接拉动经济增长。从直接方面来说，马克思的消费理论认为，消费创作劳动者主体，而劳动力的生产可以带动社会的生产。满足居民本身的消费需求可以带来一定的经济产值。从间接方面来说，居民的消费需求可以带动其他方面的需求，从而间接促进经济的发展。比如，居民的消费需求增加时，生产厂商会扩大生产规模，可能在改进生产技术方面增加投资。这样，居民消费提高拉动投资间接拉动了经济增长。

# 3 数据与模型（Data and Models）

## 3.1 变量的选择（Selection of Variables）

本文选取国内生产总值作为经济增长的度量。居民消费水平包括城镇居民消费水平和农村居民消费水平，通过城乡居民购买产品和劳务的数量和质量反映出来。城镇居民总消费是城镇居民消费水平与城镇人口的乘积，农村居民总消费是农村居民消费水平与农村人口的乘积，其反映的是城镇或农村的整个消费市场。本文对城市居民消费和农村居民消费分开讨论，以获得对比。

综上所述，本文选取了五个指标，即国内生产总值、城镇居民消费水平、农村居民消费水平、城镇居民总消费和农村居民总消费，对其进行一定的处理后，通过计量经济学的方法探索它们之间的内在联系。

## 3.2 数据的来源和处理（Source and Processing of Data）

本文选取 1978 年至 2019 年的数据作为分析的总样本。之所以没有选取 1980 年以前的数据，是因为在改革开放前中国经济比较封闭，与此后的经济运行环境差异巨大。而之所以没有选取已经公布的 2020 年与 2021 年的数据，是因为在这两年间中国经济发展和居民消费水平都受到了新型冠状病毒疫情的不同程度的影响，不宜作为样本参与分析。因此，我们研究的对象是 20 世纪八九十年代和 21 世纪一二十年代共 42 年的数据。

国内生产总值原始数据来源于国家统计局，单位是亿元人民币。为了剔除价格因素，我们再使用国内生产总值指数进行调整，得到变量 GDP。城镇居民消费水平和农村居民消费水平原始数据也来自国家统计局。同样，我们采用相关指数剔除掉价格的影响。每位城镇居民的消费水平记作 urc，每位农村居民的消费水平记作 rrc，单位是元人民币。而 urc 和 rrc 两个变量乘以每年的城镇居民人口和农村居民人口即可得到城市居民总消费和农村居民总消费，单位是亿元人民币。其中，城镇居民总消费记作 URC，农村居民总消费记作 RRC。

为了消除时间序列数据中的异方差，我们对 GDP、URC、RRC 进行自然对数变换处理，得到 LGDP、LURC、LRRC 三个变量；对 urc 和 rrc 进行同样的自然对数变换处理，得到 lurc 和 lrrc 两个变量。本文使用的数据分析软件是 EViews7.2。

LGDP 随时间的变化趋势如图 1 所示。由图 1 可知，LGDP 随时间的变化几乎是线性的。

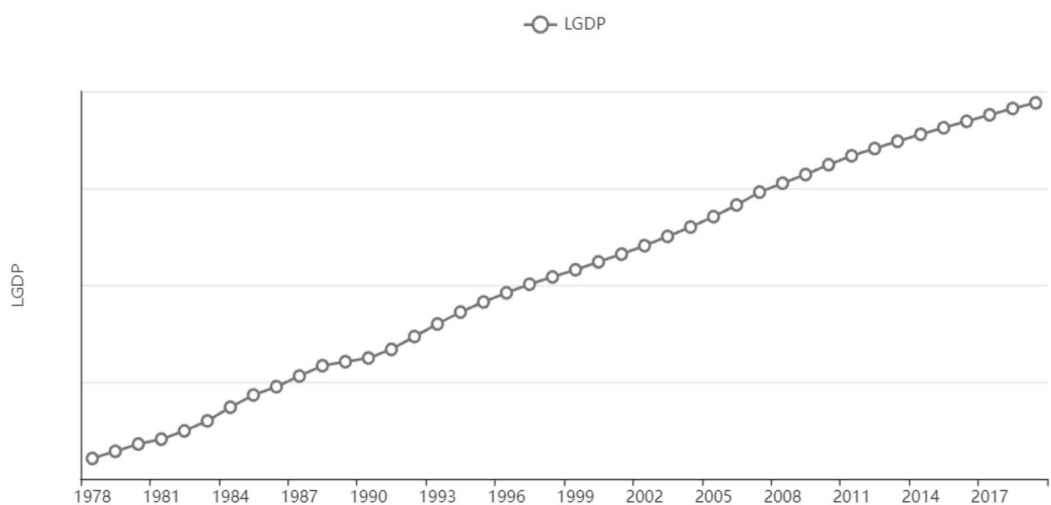

图 1　LGDP 随时间变化折线图

LURC 和 LRRC 随时间的变化趋势如图 2 所示。在 1993 年之前，由于农村人口数量众多，表

示 LURC 的折线在表示 LRRC 的折线之下；1993 年后，随着中国城镇化水平的提高，城镇居民总消费超过农村居民总消费，使得表示 LURC 的折线在表示 LRRC 的折线之上。

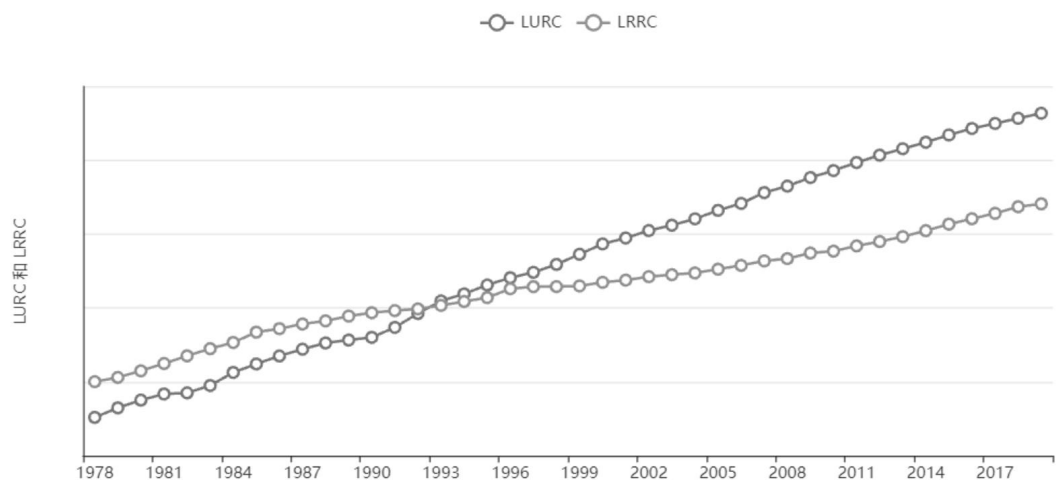

图 2　LURC 和 LRRC 随时间变化折线图

### 3.3　选用的模型（Selected Models）

VAR 模型，即向量自回归模型，是自回归模型的联立形式，可以用于描述多变量时间序列之间的变动关系。VAR 模型以多方程联立的形式出现，系统内每个方程右边变量相同，然后通过模型中所有内生当期变量对它们的若干滞后值进行回归，进而估计其动态关系。

VEC 模型，即向量误差修正模型，是有约束的 VAR 模型。VEC 模型主要约束变量的长期关系以满足其协整关系，但是允许其出现短期波动。

## 4　实证分析 A（Empirical Analysis A）

这一部分是居民消费水平与经济增长关系的实证分析，所应用到的变量是 lurc、lrrc 和 LGDP，分别表示城镇居民消费水平、农村居民消费水平和国内生产总值。

### 4.1　平稳性检验（Stationarity Test）

lurc、lrrc 和 LGDP 随时间的变化趋势如图 3 所示。横轴表示时间，纵轴表示经过自然对数变换后的数值，没有具体标出。其中上方的折线表示 LGDP，中间的折线表示 lurc，下方的折线表示 lrrc。

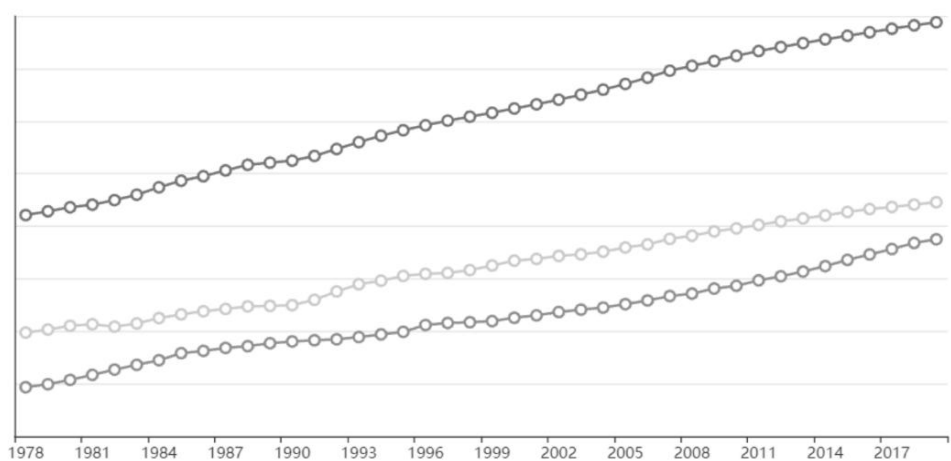

图 3　LGDP、lurc、lrrc 随时间变化折线图

容易注意到，对国内生产总值、城镇居民消费水平和农村居民消费水平进行对数变换处理后，三条折线仍然不水平，有明显的趋势变化。由此可以推断，LGDP、lurc和lrrc是非平稳的时间序列。为了验证这一猜想，可以对LGDP、lurc和lrrc三个变量进行数据平稳性检验。单位根检验（Unit Root Test）是针对宏观经济数据序列等数据序列中是否具有某种特定的统计特性而提出的一种平稳性检验的方法[4]。我们通过单位根检验判断时间序列的平稳性。ADF检验、PP检验、NP检验等多种方法都可以实现单位根检验，此处采用的是应用比较广泛的ADF检验。如果存在单位根，那么这个时间序列就是非平稳的。检验结果如表1所示。

表1　LGDP、lurc 和 lrrc 三个变量的平稳性检验结果

| 变量 | t | P | 临界值 | | |
| --- | --- | --- | --- | --- | --- |
|  |  |  | 1% | 5% | 10% |
| LGDP | -1.639 | 0.463 | -3.621 | -2.944 | -2.61 |
| lurc | -0.839 | 0.807 | -3.661 | -2.961 | -2.619 |
| lrrc | 1.583 | 0.998 | -3.601 | -2.935 | -2.606 |

该序列检验的结果显示，基于变量 LGDP，显著性 P 值为 0.463，水平上不呈现显著性，不能拒绝原假设，这说明该序列为非平稳的时间序列。基于变量 lurc，显著性 P 值为 0.807；基于变量 lrrc，显著性 P 值为 0.998。这说明以上两个序列也都是非平稳的时间序列。综上所述，猜想成立，三个序列均为非平稳的时间序列。

4.2　回归方程的建立（Establishment of Regression Equations）

我们首先尝试通过最小二乘法对 lurc、lrrc 和 LGDP 进行线性回归拟合。得到的结果如表 2 所示。

表2　lurc、lrrc 和 LGDP 的线性回归拟合结果

|  | 非标准化系数 | | 标准化系数 | t | P | VIF | $R^2$ | F |
| --- | --- | --- | --- | --- | --- | --- | --- | --- |
|  | B | 标准误 | Beta |  |  |  |  |  |
| 常数 | -0.229 | 0.106 |  | -2.167 | 0.036** |  | 0.997 | 6148.958(0.000***) |
| lurc | 1.253 | 0.075 | 0.861 | 16.801 | 0.000*** | 32.411 |  |  |
| lrrc | 0.209 | 0.077 | 0.139 | 2.715 | 0.010*** | 32.411 |  |  |

注：***、**、*分别代表 1%、5%、10%的显著性水平。

得到回归方程如下：

$$LGDP = -0.229 + 1.253 * lurc + 0.209 * lrrc$$

但是，变量 lrrc、lurc 的 VIF 值为 32.411 大于 10，存在共线关系。在此条件下，建立岭回归模型能得到更加准确的结果。于是我们尝试建立岭回归模型。

表3　lurc、lrrc 和 LGDP 的岭回归拟合结果

|  | 非标准化系数 | | 标准化系数 | t | P | $R^2$ | F |
| --- | --- | --- | --- | --- | --- | --- | --- |
|  | B | 标准误 | Beta |  |  |  |  |
| 常数 | -0.001 | 0.106 |  | -0.007 | 0.995 | 0.996 | 4897.297 (0.000***) |
| lurc | 1.02 | 0.048 | 0.701 | 21.456 | 0.000*** |  |  |
| lrrc | 0.441 | 0.049 | 0.293 | 8.982 | 0.000*** |  |  |

注：***、**、*分别代表 1%、5%、10%的显著性水平。

上表中岭回归的结果显示，基于 F 检验的显著性 P 值为 0.000，水平上呈现显著性，表明自变量 lurc 和 lrrc 与因变量 LGDP 之间存在着回归关系。同时，模型的拟合优度 $R^2$ 为 0.996，模型表现

较为优秀。岭回归方程为：

$$LGDP = -0.001 + 1.02 * lurc + 0.441 * lrrc$$

以上方程的适用范围是给定数据的时间及其前后一段时间。以上岭回归方程说明了 lurc、lrrc 和 LGDP 之间的关系：lurc 与 LGDP 呈正相关，lrrc 与 LGDP 也呈正相关，具有相同的变化趋势，且 lurc 的变化对 LGDP 的影响大于 lurc 的变化对 LGDP 的影响。定量来说的话，城镇居民消费水平每提高 1 个百分点，经济发展水平提高 1.02 个百分点；城镇居民消费水平每提高 1 个百分点，经济发展水平提高 0.441 个百分点。

### 4.3　Johansen 协整检验（Johansen Cointegration Test）

据图可观察到，三条折线的变化趋势几乎同步。因此我们可以猜测，三个变量具有协整关系。我们可以使用 Johansen 协整检验来验证这一猜测。下面是输出的结果。

表 4　lurc、lrrc 和 LGDP 的 Johansen 协整检验结果

| 原假设 | 特征根 | 迹统计量 | 5%临界值 | P |
|---|---|---|---|---|
| 无协整向量* | 0.955945 | 146.1710 | 29.79707 | 0.0001 |
| 至多一个协整向量* | 0.688411 | 43.13416 | 15.49471 | 0.0000 |
| 至多两个协整向量* | 0.131532 | 4.653819 | 3.841466 | 0.0310 |

注：*表示在 5%显著水平上拒绝原假设。

对模型进行 Johansen 协整检验的结果表明国内生产总值、城镇居民消费水平、农村居民消费水平三者具有长期的协整关系。因为 LGDP、lurc 和 lrrc 三个变量通过了协整检验，存在长期的均衡关系，所以我们可以根据 VEC 模型用每个变量对它本身和其他变量的滞后项以及误差修正项的滞后项做回归。

然后，我们使用 EViews 软件进行 VEC 模型构建。

## 5　实证分析 B（Empirical Analysis B）

这一部分是居民总消费与经济增长关系的实证分析,所应用到的变量是 LURC、LRRC 和 LGDP，分别表示城镇居民总消费、农村居民总消费和国内生产总值。

### 5.1　平稳性检验（Stationarity Test）

在时间间距相等的条件下，用后一个数值减去前一个数值，就可以得到一阶差分。在一阶差分的基础上再用后一个数值再减去前一个数值一次，就可以得到二阶差分。通过 ADF 检验结果，分析判断其是否可以显著地拒绝序列不平稳的原假设。若呈显著性，表明拒绝序列不平稳的原假设，该序列为一个平稳的时间序列。若不呈显著性，则接受原假设，说明该序列为不平稳的时间序列。不呈现显著性时，我们可以考虑对数据进行差分后继续检验，但考虑到其经济学方面的实际意义，一般不超过二阶差分。

表 5　变量 LGDP 的 ADF 检验结果

| 变量 | 差分阶数 | t | P | AIC | 临界值 | | |
|---|---|---|---|---|---|---|---|
| | | | | | 1% | 5% | 10% |
| LGDP | 0 | -1.765 | 0.398 | -171.836 | -3.621 | -2.944 | -2.61 |
| | 1 | -3.835 | 0.003*** | -164.434 | -3.621 | -2.944 | -2.61 |
| | 2 | -5.16 | 0.000*** | -162.218 | -3.621 | -2.944 | -2.61 |

注：***、**、*分别代表 1%、5%、10%的显著性水平。

表 6 变量 LURC 的 ADF 检验结果

| 变量 | 差分阶数 | t | P | AIC | 临界值 | | |
|---|---|---|---|---|---|---|---|
| | | | | | 1% | 5% | 10% |
| LURC | 0 | -3.723 | 0.004*** | -138.934 | -3.679 | -2.968 | -2.623 |
| | 1 | -4.106 | 0.001*** | -119.958 | -3.616 | -2.941 | -2.609 |
| | 2 | -7.541 | 0.000*** | -127.237 | -3.654 | -2.957 | -2.618 |

注：***、**、*分别代表 1%、5%、10%的显著性水平。

表 7 变量 LRRC 的 ADF 检验结果

| 变量 | 差分阶数 | t | P | AIC | 临界值 | | |
|---|---|---|---|---|---|---|---|
| | | | | | 1% | 5% | 10% |
| LRRC | 0 | -1.188 | 0.679 | -130.59 | -3.61 | -2.939 | -2.608 |
| | 1 | -4.205 | 0.001*** | -122.392 | -3.616 | -2.941 | -2.609 |
| | 2 | -2.149 | 0.225 | -116.953 | -3.7 | -2.976 | -2.628 |

注：***、**、*分别代表 1%、5%、10%的显著性水平。

上述 ADF 检验结果说明，变量 LGDP、LURC 的时间序列在不差分、一阶差分和二阶差分时，水平上都呈现显著性，都是平稳的时间序列；变量 LRRC 的时间序列在一阶差分时，水平上呈现显著性，是平稳的时间序列。因此，可以选择一阶差分后的 LGDP、LURC 和 LRRC 进行后续操作。

## 5.2 滞后阶数的选择（Choice of Lag Order）

表 8 滞后期评估检验结果

| 滞后阶数 | LogL | LR | FPE | AIC | SC | HQ |
|---|---|---|---|---|---|---|
| 0 | 215.0084 | NA | 2.30e-10 | -13.67796 | -13.53919 | -13.63273 |
| 1 | 233.5033 | 32.21690* | 1.25e-10 | -14.29054 | -13.73545* | -14.10959 |
| 2 | 242.0441 | 13.22446 | 1.32e-10 | -14.26091 | -13.28950 | -13.94426 |
| 3 | 249.9213 | 10.67225 | 1.49e-10 | -14.18847 | -12.80074 | -13.73610 |
| 4 | 263.0257 | 15.21804 | 1.25e-10 | -14.45327 | -12.64922 | -13.86519 |
| 5 | 273.2954 | 9.938451 | 1.36e-10 | -14.53519 | -12.31482 | -13.81140 |
| 6 | 280.7031 | 5.735022 | 1.98e-10 | -14.43246 | -11.79577 | -13.57297 |
| 7 | 306.3246 | 14.87700 | 1.07e-10 | -15.50482 | -12.45181 | -14.50961 |
| 8 | 340.1604 | 13.09771 | 4.81e-11* | -17.10712* | -13.63780 | -15.97621* |

*表示该标准下的最佳滞后期。

LR、FPE、AIC、SC、HQ 是多个评估滞后期优劣与否的指标。星号表示某一标准下的最佳滞后期，带星号最多的滞后期即为我们要选择的最佳滞后期。由表中 LR、FPE、AIC、SC、HQ 等多项评价指标的结果，滞后阶数可以选为 8 阶。因此，我们可以建立 VAR(8)模型。

## 5.3 VAR 模型的建立（Establishment of VAR Model）

根据上述分析，我们可以根据一阶差分后的数据建立VAR(8)模型。通过EViews软件对数据进行分析，得到对应的系数。

## 5.4 脉冲响应分析（Impulse Response Analysis）

脉冲响应分析可以用来描述一个内生变量对另一个内生变量所带来的冲击的反应，即在一个内生变量上施加一个标准差大小的冲击后，对另一个内生变量的当期值和未来值所产生的影响程度。

由脉冲分析结果可知，D(LGDP)对其自身冲击的响应在第 3 期后变换正负，对 D(LURC)冲击的响应的整体幅度较大，并在第 7 期后趋于稳定，对 D(LRRC)冲击的响应整体幅度较小。

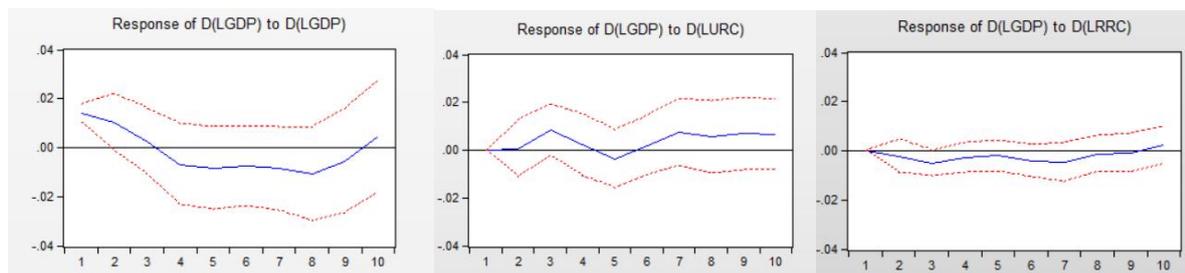

图 4　D(LGDP)受到 D(LGDP)、D(LURC)和 D(LRRC)冲击的脉冲响应分析图

### 5.5　方差分解（Variance decomposition）

方差分解是把系统中内生变量的波动按其成因分解为多个组成部分。由方差分解图可知，D(LGDP)的变化在初期主要是由于其自身，在第 1 期的贡献率几乎达到了 100%。在第 2 期和第 3 期，D(LURC)、D(LRRC)对 D(LGDP)的变化的贡献率提高，此后基本达到稳定。

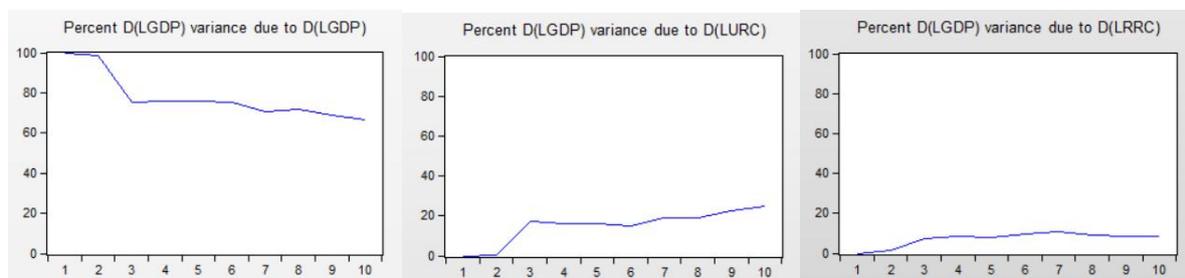

图 5　D(LGDP)的方差分解图

## 6　结论及原因分析（Conclusion and Cause Analysis）

根据实证分析 A 得到的方程，我们可以得出结论：城镇居民消费水平和农村居民消费水平的提升都能导致国内生产总值的提高，且城镇居民消费水平对经济增长的促进能力比农村居民消费水平更强。城镇居民消费水平每提高 1 个百分点，经济发展水平提高 1.02 个百分点；城镇居民消费水平每提高 1 个百分点，经济发展水平提高 0.441 个百分点。

综合两部分实证分析的结果，我们还能知道，国内生产总值和城镇居民消费水平与农村居民消费水平具有长期均衡的关系。也就是说，经济发展和城乡居民消费具有长期均衡的关系。城镇居民消费和农村居民消费都能促进经济发展，但从长期来看，前者对经济发展的贡献率要高于后者。反过来，国内生产总值的提高能带动城镇居民消费和农村居民消费的提高，即经济发展能促进城乡居民消费。

因此，要想在长期内提高中国经济发展水平，可以提高居民消费，尤其是在城镇地区。经济发展的水平的提高再促进城镇居民消费和农村居民消费，从而形成一个良性的循环。

在消费带动经济增长方面，中国自身有着一定的优势。中国作为发展中国家，城乡差别较大，东西部地区差别较大，城镇化发展还有很大的空间。人民的生活水平、消费水平、收入水平和发达国家还有一定的差距，在这些方面中国有着巨大的潜力。除此之外，中国幅员辽阔、人口众多，拥有超大规模的内需市场，发展空间巨大。

从需求端看，城镇地区人口较为集中，对消费尤其是服务类的消费拉动作用显著。人口聚集意味着有集中的市场需求。服务业大都是直接面对客户的，比如电影院、歌剧院、体育场，许多服务

的提供和消费对供需双方的空间距离提出了较高的要求。如果供需双方空间距离过大，比如电影院、歌剧院离居民区较远，则电影和歌剧对居民的吸引力会大打折扣。再如，农村地区的外卖产业远没有城镇地区发达，一方面是因为农村居民的生活习惯与传统，另一方面则是因为农村地区居民居住较散，人口密度较低，不利于外卖产业的发展。从供给端看，由于产业的规模效应，城镇地区信息成本、物流成本、交易成本更低，经济运行效率更高。城镇化有利于服务消费的发展。

同时，我们也应该看到经济高度发达的城市有可能给周边城市带来"虹吸效应"，不利于经济社会持续有效发展。如果经济高度发达的城市对周边的虹吸效应大于外溢效应，会加剧区域间发展的不平衡，不利于城市群的建设[5]。

# 7 政策建议（Policy Suggestions）

基于本文的实证分析结果和中国的实际情况，我们提出以下政策建议。

建立健全社会保障体系，提高低收入人群及受疫情影响人群的收入。为居民提供更好的社会保障设施与服务，可以提高居民的消费保障，从而提高消费的意愿，带动消费水平和总消费的提高。中国的社会保障体系包括医疗、养老、教育、住房、就业等方面，与居民生活息息相关。这一体系还面临着覆盖率不够全面等问题，需要在探索和实践中进一步完善[6]。在新冠疫情的阴霾下，许多企业遇到了经营困难，许多居民面临着失业的风险。通过一定方法帮助他们解决当前遇到的问题，帮助他们走出困境，是一种可以刺激消费的有效援助政策。

加强乡村振兴工作，提高农村人口消费能力。2020 年，中国现行标准下贫困人口全部脱贫，贫困地区全部"摘帽"，这一伟大壮举有助于充分挖掘贫农村人口的消费潜力，带动经济的高质量发展。在此后的乡村振兴战略中，应当注意健全农村地区的社会保障体系，挖掘农村地区的消费潜力。

缩小居民收入差距，降低基尼系数。过大的城乡居民收入差距不利于经济发展，甚至有可能对经济的发展造成严重的阻碍。在过大的居民收入差距下，居民经济方面的的抗风险能力降低，消费意愿降低。在经济社会迅速发展的同时，中国城乡二元结构等历史遗留问题尚未得到充分有效的解决，这使得城乡收入差距较大，不利于社会和谐稳定。

提高城镇化水平，优化三次产业在国民经济中的比例。当前，越来越多的农村人口选择进城就业，他们的收入相比在农村从事农业活动时得到了一定的提升，消费能力也得到了增强。中国近年来的城镇化速度较快，在较短的时间内完成了一些西方国家花费更多年才完成的城镇化任务。因而，在改革开放后的这一时期内，中国经济发展取得了举世瞩目的成就，创造了彪悍史册的经济奇迹。在科技继续发展进步的未来，我们可以进一步合理提升城镇化水平，增加一次产业在国民经济中的比重，从而提高经济发展水平和居民生活质量。

调整和优化产业结构，构建现代产业体系。对于有较大发展潜力的传统产业，如中国东北地区的重金属产业、山西地区的煤矿产业等，需要对传统技术进行改造，提高劳动生产率和资源利用率。

鼓励科技创新，深化科技体制改革，增加在创新方面的投入，推动创新驱动发展战略，鼓励各企业、从业人员和高校提出创新方案。加大教育投入，加强创新人才队伍建设，培养更多的创新型人才。

# 参考文献